\documentclass{c2k}

\usepackage{graphicx}

\begin{document}
\begin{titlepage}

\title{Sun Microsystems' AutoClient \/ and management of computer farms at BaBar}

%

\author{A.V.~Telnov \inst{1,2}, S.~Luitz\inst{3},
T.J.~Pavel \inst{3}, O.H.~Saxton \inst{3}, M.R.~Simonson \inst{3}}

\institute{Lawrence Berkeley National Laboratory, USA
\and University of California at Berkeley, USA
\and Stanford Linear Accelerator Center, USA }

\maketitle


\begin{abstract}
Modern HEP experiments require immense amounts of computing power. In the 
BaBar experiment at SLAC, most of it is provided by Solaris SPARC systems. 
\emph{AutoClient}, a product of Sun Microsystems, was designed to make setting up 
and managing large numbers of Solaris systems more straightforward. 
\emph{AutoClient} machines keep all filesystems, except \emph{swap}, on a server and 
employ \emph{CacheFS} to cache them onto a local disk, which makes them Field 
Replaceable Units with performance of stand-alone systems. We began 
exploring the technology in Summer 1998, and currently operate online, 
reconstruction, analysis and console AutoClient farms with the total 
number of nodes exceeding 400. Although the technology has been available 
since 1995, it has not been widely used, and the available documentation 
does not adequately cover many important details of \emph{AutoClient} 
installation and management. This paper discusses various aspects of our 
experience with \emph{AutoClient}, including tips and tricks, performance and 
maintainability, scalability and server requirements, existing problems 
and possible future enhancements. 

{\ }

\emph{This paper has been submitted to Proceedings of the Conference on Computing in 
High Energy and Nuclear Physics (CHEP 2000), February 7-11, 2000, Padova, Italy.}

\end{abstract}


\end{titlepage}


\section{Introduction}
\label{sec:introduction}
    
Having to administer a large number of workstations has long been a headache to system administrators
working for big businesses, universities and research laboratories. In the case of UNIX-style operating systems, 
system maintenance tasks can very rarely be delegated to the end users, which means that the whole burden 
of taking care of the company's computers falls on the system administrators. Unless some kind of special system 
maintenance scheme is devised, the required 
administration effort scales linearly with the number of machines and a synchronous major software upgrade
is essentially impossible (unless, of course, the company employs an army of sysadmins and the upgrade takes place 
during the Christmas shutdown).

The \emph{Ba\={B}ar} experiment\footnote{\emph{Ba\={B}ar} homepage: \url{http://www.slac.stanford.edu/BFROOT/}.}, 
which began operation at the Stanford Linear Accelerator Center in 1999, 
requires an immense amount of computing power, most of which is currently provided by Solaris SPARC systems. 
The 300+ -node Analysis and Prompt Reconstruction farm  
--- and especially the 79-node Online Data Flow farm ---
perform tasks that are critical for experiment operation and demand minimal downtime in case of a 
hardware malfunction, operating system crash, or a software upgrade. 

Without using any special scripts or system administration tools, setting up a Sun Solaris stand-alone ``from scratch'' 
and configuring it to make the best use of the SLAC 
computing environment (AFS, NIS, AMD, etc.) and to conform to certain security standards 
(\texttt{ssh}, \texttt{sudo}, \texttt{sendmail}, disabling \texttt{telnet} and \texttt{rlogin}, applying 
patches) is a task that takes at least 10 hours, of which 3 to 5 hours require active administrator involvement.
\texttt{Tailor} --- a collection of system administration tools developed by the SLAC Computing Services
over the past 10 years for many flavors and versions of UNIX --- greatly simplifies integration of 
a UNIX workstation into the SLAC environment and saves a lot of time: the process takes 3 to 5 hours
with periodical administrator involvement totalling about 1 hour per machine. 

But even with the help of
\texttt{tailor}, managing such a large number of machines is a formidable task, so in Summer 1998 we started 
looking for ways to make deployment or replacement of Solaris machines faster and system management more straightforward. 
The three techniques that we have come up with are 1) ``non-invasive'' hard drive cloning
with the help of the \emph{Diskless Client} technology\footnote{The 
process is described in detail in \cite{babarnote}, section 5.}, which allows 
us to make a fully functional copy of a `\texttt{tailor}ed' Solaris stand-alone system in about 20 minutes 
(including changing identity information); 2) A combination of \texttt{tailor} and Sun's 
\emph{JumpStart}\footnote{The \emph{JumpStart} technology is described in the \emph{SPARC: Installing
Solaris Software} manual from Sun Microsystems.}, 
which requires a network installation server with custom install/finish scripts that allow \texttt{tailor}ing
to take place without administrator intervention (this technique of making a Solaris stand-alone takes about
1 hour plus 30 to 60 minutes to initialize the AFS cache); and 3) Sun Microsystems' \emph{AutoClient}, which 
will be discussed in the rest of this article.


\section[Overview of {\em AutoClient}]{Overview of {\em AutoClient}\/\footnote{For a more detailed discussion, 
please refer to \cite{babarnote} and \cite{babarweb}.}}
\label{sec:overview}

Sun Microsystems \emph{AutoClient} and \emph{AdminSuite} products were designed to centralize and
simplify administration of a large number of Solaris workstations. To understand what an AutoClient is, let's
compare it with a stand-alone system and a diskless client:

 \begin{table}[ht]
 \begin{center}
 \caption{\emph{AutoClient} vs. \emph{Diskless Client} and stand-alone systems}
 	\begin{tabular}{|p{1.8cm}|p{3.2cm}|p{0.9cm}|p{3.5cm}|p{1.1cm}|p{1.7cm}|}
	 \hline
 	 \textsf{\textbf{\scriptsize {\raggedright System Type}}} & 
		\textsf{\textbf{\scriptsize {\raggedright Local File Systems}}} & 
		\textsf{\textbf{\scriptsize {\raggedright Local \\ Swap?}}} & 
		\textsf{\textbf{\scriptsize {\raggedright Remote File Systems}}} & 
		\textsf{\textbf{\scriptsize {\raggedright Network \\ Use}}} & 
		\textsf{\textbf{\scriptsize {\raggedright Relative \\ Performance}}} \\ 
	\hline
 	\textsf{\textsl{\scriptsize {\raggedright Stand-alone \\ System}}} & 
		\textsf{\scriptsize {\raggedright root (\texttt{/}), \texttt{/usr}, 
		\texttt{/opt},\\ \texttt{/export/home}}} & \textsf{\scriptsize Yes} & 
		\textsf{\scriptsize {\raggedright -none- (meaning ``not \\ necessary'')}} & 
		\textsf{\scriptsize Low} & \textsf{\scriptsize High}  \\
 	\hline
 	\textsf{\textsl{\scriptsize {\raggedright Diskless Client}}} & \textsf{\scriptsize -none-} & 
		\textsf{\scriptsize No} & 
		\textsf{\scriptsize {\raggedright root (\texttt{/}), \texttt{swap}, \texttt{/usr}, \texttt{/opt}, 
		\\ \texttt{/home}}} & 
		\textsf{\scriptsize High} & \textsf{\scriptsize Low}  \\
	\hline
 	\textsf{\textsl{\scriptsize {\raggedright AutoClient \\ System}}} & 
		\textsf{\scriptsize {\raggedright cached root (\texttt{/}), cached \\ \texttt{/usr}, 
		cached \texttt{/opt}}} & \textsf{\scriptsize Yes} & 
		\textsf{\scriptsize root (\texttt{/}), \texttt{/usr}, \texttt{/opt}, \texttt{/home}} & 
		\textsf{\scriptsize Low} & \textsf{\scriptsize High}  \\
	\hline	
 \end{tabular}
 \label{the table}
 \end{center}
 \end{table}

\emph{AutoClient} allows us to keep all clients' file systems, except \texttt{swap}, on the AutoClient Server
and \emph{\textbf{locally cache}} root (\texttt{/}) and the shared read-only \texttt{/usr} and \texttt{/opt} using 
the \emph{CacheFS} technology, which is the most important component of \emph{AutoClient}. 

\emph{CacheFS} caches
files that have been accessed by the AutoClient, so that subsequent requests to the same files get referenced to the
cache rather than being sent to the server. A cache consistency check is performed every 24 hours by a cron job 
running on the AutoClient, on reboot, 
or at request. All writes immediately update the back file system on the server, unless the AutoClient is configured
as `disconnectable' and the server is temporarily unavailable. This consistency check policy relies on the assumption 
that the cached file systems do not get changed from the server side except in rare cases by the administrator who 
explicitly requests a consistency check after he is done. Specific files or directories can be \emph{\textbf{packed}} into the 
\emph{CacheFS} cache, which guarantees that they will always be in the cache and will not be purged if the cache becomes
full. This feature can be particularly useful with `disconnectable' clients.

In a nutshell, an \emph{AutoClient} system has all the advantages of a diskless
client (with the exception of not needing a hard drive) while not putting a heavy load on the network and
possessing performance closely matching that of a stand-alone system.

Since AutoClients do not require any swap space on the server and share \texttt{/usr} and \texttt{/opt} 
filesystems\footnote {The AutoClient Server can also be configured to serve its own \texttt{/usr} and \texttt{/opt} 
to the AutoClients.}, each AutoClient requires only about 40 MB of space on the server for its root file system.
In order to backup each \emph{AutoClient} system, we only need to backup 
the server. We also can manipulate AutoClient root file systems (read log files, apply patches, etc.) directly from the server. 
AutoClients can be configured to be \emph{\textbf{disconnectable}}, which means that they will continue to function using their cached 
filesystems while the server is temporarily unavailable. AutoClients can be halted and rebooted remotely; they reboot directly
from the cache, so the network traffic during a system-wide reboot is limited to a cache consistency check. 
Since no persistent information is stored on the AutoClient itself, 
it can be considered a \emph{field-replaceable unit \textbf{(FRU)}}. Replacing a failed unit or deploying a new AutoClient takes just a 
few minutes\footnote{Plus, if AFS is used, the time required to build a new local AFS cache.}. Most of the management tasks
normally associated with stand-alone Solaris systems are thus almost completely eliminated.

The quintessence of the centralized administration model, of which \emph{AutoClient} is a key component, 
is a significant reduction of the cost of management --- that is, if everything works as advertised.


\section{Ba\={B}ar's experience with \emph{AutoClient}}
\label{sec:babarexperience}

We started experimenting with \emph{AutoClient} in June 1998, about a year before \emph{Ba\={B}ar} took its first 
$e^{+}e^{-}$ collision data. Our
first AutoClient server and two dozen or so AutoClients were Ultra-5's with a 270 MHz UltraSPARC-IIi CPU, 128 MB RAM and a 
4.3 GB 5,400 rpm EIDE HDD running
Solaris 2.6 HW 3/98; we used this prototype AutoClient farm as console and Online Data Flow machines during the Winter 1998/99 
\emph{Ba\={B}ar}
cosmic ray run. In order to speed up creation of additional AutoClients, we developed a set of scripts that `clone' the root
file system of a fully configured AutoClient, modify identity-related files in \texttt{/export/root/\textsl{clientname}}, 
and make necessary adjustments to the server configuration files --- all in about one minute.

Although the process of configuring the server and the first fully functional client was quite bumpy 
(mostly having to do with getting proper patches installed, see \cite{babarnote}), we were satisfied with the farm's performance 
and decided to use the \emph{AutoClient} technology on all \emph{Ba\={B}ar} computer farms at SLAC. At this time (January 2000), 
there are 309 AutoClients\footnote{Currently, 
Ultra 5's with a 333 MHz UltraSPARC-IIi CPU, a 9.1 GB EIDE HDD and 256 MB RAM, soon to be replaced with
rack-mountable Netra t1's with a 440 MHz UltraSPARC IIi CPU, two 9.0 GB 10,000 rpm SCSI HDDs and 256 MB RAM.} 
on 6 AutoClient Servers\footnote{Ultra 2's with two 296 MHz UltraSPARC II CPUs and 9.0 GB SCSI HDDs.
} in the Analysis and Prompt Reconstruction farm, which is located in the SCS building, and 100 AutoClients\footnote{Mostly 
Ultra 5's with a 333 MHz UltraSPARC-IIi CPU, a 9.1 GB EIDE HDD and 512 MB RAM.} and 1 AutoClient Server\footnote{An 
Enterprise 450 server with
four 296 MHz UltraSPARC-II CPUs, 2 GB RAM, two 4.2 GB SCSI HDDs and four 188 GB Baydel RAID Level 3 arrays.} in the Online Data Flow and console farms, 
which are located in the IR-2 building that houses the \emph{Ba\={B}ar} detector. 

So, we have been operating over 400 AutoClients under real life conditions (running online, prompt reconstruction and analysis jobs 
around the clock at close to 100\% capacity) for about 8 months. Overall, the farms performed their goals very well. However, 
the required management effort turned out to be much bigger than we expected, primarily because of a bug in \emph{CacheFS}, which
has been identified by Sun Microsystems. A fix for this bug is reportedly available for Solaris 7, but Sun still has not been 
able to come up with a fix for Solaris 2.6 that \emph{Ba\={B}ar} is currently using. 

The bug leads to cache corruption and, occasionally, to
disappearance of files on the client's root file system during power outages or if connection to the server is lost due to a 
server reboot or crash or a network outage, probably only if the AutoClient was in the process of writing into a file.
Such accidents have so far occurred about one a week, and each time about 15-20\% of 
AutoClients had to be manually rebooted with the \texttt{boot -f} command that forces cache reconstruction; sometimes
an AutoClient had to be recloned. This means that after an outage the status of each AutoClient has to be checked manually or 
\emph{all} AutoClients have to be rebooted with \texttt{boot -f} --- either way, this takes a lot of time. 

It also turns out that while in most cases a Solaris stand-alone system does not have to be rebooted after a patch is applied to it, 
AutoClients often do, the reason being differences in the UFS and NFS file locking mechanisms. An \emph{AutoClient} system 
has to be idle before patching takes place --- otherwise running applications can crash; 
a global farm outage has to be scheduled to patch \texttt{/usr}.  

We have undertaken several measures to minimize the impact of the outages: the AutoClient Servers and network equipment at SCS have
been connected to UPS power; the network topology has been modified to remove path redundancies that under certain circumstances can 
lead to brief periods of network unavailability. We have also realized that putting the responsibility of being the AutoClient Server
on the main IR-2 server was a big mistake because it often crashed due a kernel memory leak or had to be rebooted. 

\subsection{Conclusion}

At this point, we are quite disappointed by our experience with \emph{AutoClient}, and unless Sun fixes the \emph{CacheFS} bug in the
nearest future, we will replace AutoClients in the SCS farm with Solaris stand-alone systems which will be net-booted from 
a net-install server using \emph{JumpStart} and \texttt{tailor}; the recovery strategy in this case would be to reinstall. We are
far from certain whether we will completely drop the \emph{AutoClient} technology and think that it has a great potential, so we 
want to try the more classical approach and see how the management effort compares to using \emph{AutoClient}. The main goal 
of our presentation at CHEP 2000 was to make the High Energy Physics community aware of \emph{AutoClient}'s existence, its pros and
cons, and our experience with it --- and let you decide whether you want to try it out or not. We hope that this goal has been achieved.



\end{document}